\documentclass[aps,pra,preprint,amsmath,amssymb]{revtex4}
\usepackage{graphicx,epsfig,epsf}
\usepackage{dcolumn}
\usepackage{bm,psfrag,color}
\usepackage[matrix,frame,arrow]{xypic}

\begin{document}

\newcommand{\ri}{{\rm i}}
\newcommand{\re}{{\rm e}}
\newcommand{\bx}{{\bf x}}
\newcommand{\bd}{{\bf d}}
\newcommand{\br}{{\bf r}}
\newcommand{\bk}{{\bf k}}
\newcommand{\bE}{{\bf E}}
\newcommand{\bR}{{\bf R}}
\newcommand{\bM}{{\bf M}}
\newcommand{\bn}{{\bf n}}
\newcommand{\bs}{{\bf s}}
\newcommand{\tr}{{\rm tr}}
\newcommand{\tbs}{\tilde{\bf s}}
\newcommand{\rSi}{{\rm Si}}
\newcommand{\beps}{\mbox{\boldmath{$\epsilon$}}}
\newcommand{\bthe}{\mbox{\boldmath{$\theta$}}}
\newcommand{\rg}{{\rm g}}
\newcommand{\xmax}{x_{\rm max}}
\newcommand{\ra}{{\rm a}}
\newcommand{\rx}{{\rm x}}
\newcommand{\rs}{{\rm s}}
\newcommand{\rP}{{\rm P}}
\newcommand{\up}{\uparrow}
\newcommand{\down}{\downarrow}
\newcommand{\hc}{H_{\rm cond}}
\newcommand{\kb}{k_{\rm B}}
\newcommand{\cI}{{\cal I}}
\newcommand{\tit}{\tilde{t}}
\newcommand{\cE}{{\cal E}}
\newcommand{\cC}{{\cal C}}
\newcommand{\Ubs}{U_{\rm BS}}
\newcommand{\qq}{{\bf ???}}
\newcommand*{\etal}{\textit{et al.}}

\sloppy

\title{Decoherence-enhanced measurements}
\author{Daniel Braun$^{1,2}$ and John Martin$^{1,2,3}$}

\affiliation{$^{1}$Universit\'e de Toulouse, UPS, Laboratoire
de Physique Th\'eorique (IRSAMC), F-31062 Toulouse, France}
\affiliation{$^{2}$CNRS, LPT (IRSAMC), F-31062 Toulouse, France}
\affiliation{$^{3}$Institut de Physique Nucl\'eaire, Atomique et de
Spectroscopie, Universit\'e de Li\`ege, 4000 Li\`ege, Belgium}

\maketitle



{\bf Quantum-enhanced measurements use highly non-classical quantum
  states in order to enhance the sensitivity of the measurement of
  classical quantities, like the length of an optical cavity
  \cite{Giovannetti04}. The major goal is to beat the standard quantum
  limit (SQL), i.e.~a sensitivity of order $1/\sqrt{N}$, where $N$ is
  the number of quantum resources (e.g.~the number of photons or atoms
  used), and to achieve a scaling $1/N$, known as the Heisenberg
  limit. Doing so would have tremendous impact in many areas
  \cite{Huelga97,Goda08,Budker07}, but so far very few experiments
  have demonstrated a slight improvement over the SQL
  \cite{Leibfried05,Nagata07,Higgins07}. The required quantum states
  are generally difficult to produce, and very prone to
  decoherence. Here we show that decoherence itself may be used as an
  extremely sensitive probe of system properties. This should allow
  for a new measurement principle with the potential to achieve the
  Heisenberg limit without the need to produce highly entangled
  states.}

Decoherence arises when a quantum system interacts with an environment with
  many uncontrolled  degrees of freedom, such as the modes of the
  electromagnetic field,
  phonons in a solid, or simply a measurement
  instrument \cite{Zurek91}. Decoherence destroys quantum mechanical
  interference,
  and plays an important role in the transition from quantum to classical
  mechanics
  \cite{Giulini96}. It becomes extremely fast if the ``distance'' between
  the components of a ``Schr\"odinger cat''-type superposition of quantum
  states reaches mesoscopic or even macroscopic proportions. Universal
  power laws rule the scaling of the decoherence rates in this regime
  \cite{Braun01} and
  lead to time scales so small that in fact the founding fathers of quantum
  mechanics postulated an instantaneous collapse of the wave-function during
  measurement. Only recently could the collapse be time-resolved in
  experiments with relatively small ``Schr\"odinger cat''--states
  \cite{Brune96,Guerlin07}.  However, different superpositions may
  decohere with very different rates. In particular, if the coupling of the
  quantum system to the environment enjoys a certain symmetry, entire
  decoherence-free subspaces (DFS) may exist, in which superpositions of
  states retain their coherence, regardless of the ``distance'' between the
  superposed states. In essence, the symmetry prevents the
  environment to distinguish the states, such that no information leaks out
  of the system and the quantum superpositions remain intact. DFS have found
  widespread use in quantum
  information theory after their formulation for Markovian master
  equations \cite{Zanardi97,Lidar98,Duan98}, experimental
  demonstration \cite{Kwiat00,Kielpinski01,Viola01}, and once it was
  realized that
  quantum computation might be performed inside a DFS \cite{Beige00}.
Given the reliance of the DFS on a symmetry in the coupling to the
  environment, it is  clear
  that for a large ``Schr\"odinger cat''-type superposition
  prepared in a DFS, the decoherence rate should be extremely sensitive to
  any changes that modify the symmetry of the coupling. This is the basic
  idea underlying the new measurement principle which we call
  ``Decoherence-Enhanced Measurements'' (DEM). Two fortunate circumstances
  make us believe that this idea may be turned into something of practical
  relevance. First, while it may seem that DEMs
  would again require the extremely difficult initial preparation of
  a highly entangled macroscopic state, surprisingly  the
  Heisenberg limit can be reached with a much simpler to prepare product state of pairs of atoms. Second, the required initial
  ``symmetry'' of the coupling to the
  environment means nothing more but a degenerate eigenvalue of the
  Lindblad operators in the master equation, or, more generally, of the
  coupling Hamiltonian of the system to the environment
  \cite{Lidar98}. Actual
  symmetries in the system (e.g.~spatial symmetries) may lead to
  such degeneracy, but are by no means
  necessary \cite{Braun01B}.
  The scheme is therefore much more general than it may appear at first
  sight.

In order to illustrate the concept, we consider
$N$ two--level atoms or ions ($N$ even, ground and excited states
$|0\rangle_i$,
$|1\rangle_i$ for atom $i$, $i=1,\ldots,N$) localized in a cavity with one
semi-reflecting mirror, and
resonantly coupled with coupling constants $g_i$ to a single e.m.~mode of
the cavity of
frequency $\omega$. The reduced density matrix $\rho$ of the atoms evolves
according to the master equation in
the interaction picture
\begin{equation} \label{L}
\frac{d}{dt}\rho(t)={\cal L}_s[\rho(t)]+{\cal L}_c[\rho(t)]\,,
\end{equation}
where
\begin{equation} \label{Ls}
{\cal
  L}_s[\rho(t)]=\frac{\Gamma}{2}\sum_{i=1}^N\left(\big[\sigma_-^{(i)}\rho(t),\sigma_+^{(i)}\big]
  +\big[\sigma_-^{(i)},\rho(t)\sigma_+^{(i)}\big]\right)
\end{equation}
  describes individual spontaneous emission with rate $\Gamma$,
  while
\begin{equation} \label{Lc}
{\cal  L}_c[\rho(t)]=\gamma\,([J_-\rho(t),J_+]+[J_-,\rho(t) J_+])
\end{equation}
models collective decoherence,
$J_-=\sum_{i=1}^N\tilde{g}_i\sigma_-^{(i)}$, $J_+=J_-^\dagger$,
$\sigma^{(i)}_+=|1\rangle_{i}\langle 0|_{i}$
($\sigma^{(i)}_-=|0\rangle_{i}\langle 1|_{i}$), and
$\tilde{g}_i=g_i/g$ where $g$ is the average coupling strength over
all atoms coupled to the cavity field.  The rate $\gamma=g^2/\kappa$
(with $\kappa$ the single photon cavity decay rate) is independent of $N$.
Equation~(\ref{L}) is a well--known and experimentally verified
\cite{Gross76,Skribanowitz73} master equation which for
$\tilde{g}_i=1\,\,\forall i$ and in the bad cavity limit $\Gamma \ll
g\sqrt{N}\ll \kappa$ describes superradiance
\cite{Agarwal70,Bonifacio71a,Glauber76,Gross82}. Due to the spatial
envelope of the e.m.~mode in resonance with the atoms, the $g_i$
depend on the position $x_i$ of the atoms along the cavity axis and
on the length $L$ of the cavity (the waist of the mode is taken to
be much larger than the size of the atomic ensemble),
\begin{equation} \label{gi}
g_i=\sqrt{\frac{\hbar\omega}{\epsilon_0 V}}\sin(k_x
x_i)\,\beps\cdot\bd\,,
\end{equation}
where $k_x=\pi n_x/L$,
$\epsilon_0$  denotes the dielectric constant of vacuum, $V=LA$ the mode
volume (with an effective cross--section $A$), $\beps$ the
polarization vector of the mode, and $\bd$ the vector of electric dipole
transition matrix elements between the states $|0\rangle_i$ and
$|1\rangle_i$, taken identical for all atoms.
Decoherence in this
system has been extensively studied, see \cite{Braun01B} for a
review.
The initial state
$\rho_0=|\psi_0\rangle\langle\psi_0|$ belongs to the DFS with respect to
collective emission, ${\cal L}_c[\rho_0]=0$, if and only if
$J_-|\psi_0\rangle=0$ \cite{Lidar98}.  If $\tilde{g}_i=1$ for $i=1,\ldots,N$,
this DFS
is well known \cite{Beige00b,Braun01B}. It contains ${N \choose N/2} \sim
2^{N}/\sqrt{N}$ DF
states, including a $2^{N/2}$ dimensional subspace
$\bigotimes_{l=1}^{N/2}\{|t_-\rangle_l,|s\rangle_l\}$ in which the pair formed by the atoms $l$ and $l+N/2$  can
be in a superposition of the triplet ground state
$|t_-\rangle_l=|0\rangle_l|0\rangle_{l+N/2}$ and the singlet
$|s\rangle_l=\frac{1}{\sqrt{2}}(|0\rangle_l|1\rangle_{l+N/2}-
  |1\rangle_l|0\rangle_{l+N/2})$. For arbitrary $\tilde{g}_i$, $|s\rangle_l$
  should be replaced by $\frac{1}{\sqrt{|\tilde{g}_l|^2+|\tilde{g}_{l+N/2}|^2}}(\tilde{g}_l|0\rangle_l|1\rangle_{l+N/2}-\tilde{g}_{l+N/2}
  |1\rangle_l|0\rangle_{l+N/2})$.

\begin{figure}
\epsfig{file=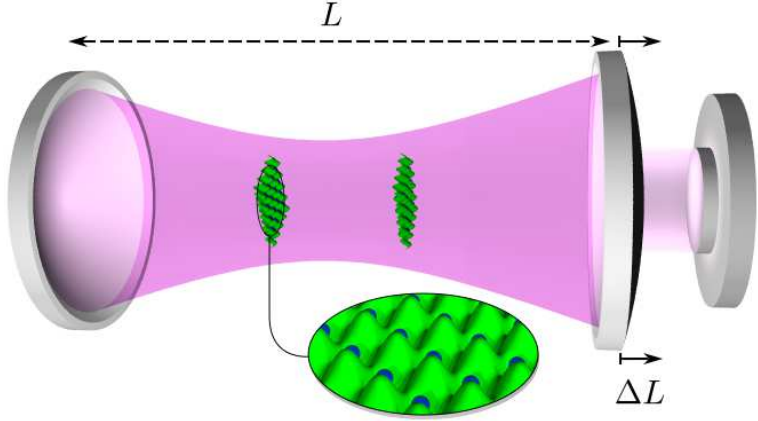,width=10cm,angle=0}\hspace{0.3cm} \caption{$N$
atoms (or ions) are trapped at
  fixed positions in two 2D optical lattices perpendicular to the cavity
  axis. A dipole transition of the atoms is in resonance with a single,
  leaky  cavity mode. The atoms are initially prepared in a DFS state
  relative to a given cavity length.  When the cavity length changes
  slightly, the DFS evolves, and the initial state is exposed to collective
  decoherence, detectable by photons leaking out through the semi-reflecting
  mirror at a rate proportional to $N^2$.
  }\label{fig.system}
\end{figure}

Consider now the
situation where the atoms
can be grouped into {\em two sets} with $N/2$ atoms each and
coupling constants $G_1$ in the first set ($i\in S_1\equiv\{1\ldots,N/2\}$),
and $G_2$ in the second
set ($i\in S_2\equiv\{N/2+1,\ldots,N\}$).
One way of obtaining two coupling constants may be to trap
the atoms in
two two--dimensional lattices perpendicular to the
cavity  axis (see Fig.\ref{fig.system}).
Suppose that after preparing the atoms in a DFS state corresponding
to the initial couplings $\tilde{G}_I^{(0)}=G_I^{(0)}/g$ ($I=1,2$),
the length $L$ of the cavity changes slightly. The coupling
constants will evolve,  $\tilde{G}_I^{(0)}\rightarrow
\tilde{G}_I\equiv G_I/g$, and  so will the DFS. It is this
collective change of the coupling constants which can be revealed
very sensitively through the decoherence it induces as the original
state becomes exposed to decoherence. The induced decoherence
therefore provides for a very precise measurement of the change of
the length of the cavity, as we shall show now.

In order to simplify notation we will assume in the following
$\tilde{G}_1^{(0)}=\tilde{G}_2^{(0)}$ (i.e.~the atoms are located for instance
symmetrically with respect to an antinode of the cavity mode, or at a
distance given by an integer multiple of the wavelength of the mode), but we
emphasize that everything goes
through for different initial couplings, unless otherwise mentioned.
Assume that an initial pure {\em product  state} of pairs of
atoms is prepared in the initial DFS,  $\rho_0=|\psi_0\rangle\langle
\psi_0|$, where
\begin{equation} \label{psi0}
|\psi_0\rangle=\bigotimes_{l=1}^{N/2}|\varphi_l\rangle_l
\end{equation}
with $|\varphi_l\rangle_l=a_l|t_-\rangle_l+b_l|s\rangle_l$. The
decoherence mechanism (\ref{Lc}) is directly linked to photon loss
from the cavity.  The induced decoherence can be measured through
the number of photons $n_{\rm ph}$ which escape through the cavity
mirror during a small time interval $\Delta t$. In the superradiant
regime considered here ($\Gamma \ll g\sqrt{N}\ll \kappa$), any
photon created leaves the cavity immediately, such that the quantum
expectation value of $n_{\rm ph}$  is given by $\langle n_{\rm
ph}\rangle=-\langle\dot{J}_z(t=0)\rangle_c\Delta
t=-\tr\big(J_z\mathcal{L}_c[\rho_0]\big)\Delta t$, where the
collective pseudo-spin component
$J_z=\frac{1}{2}\sum_{i=1}^N\sigma_{z}^{(i)}$ measures total
population inversion of the atoms. As long as $|\psi_0\rangle$
resides in the DFS, we have $\langle n_{\rm ph}\rangle=0$. If the
coupling constants undergo slight changes and get replaced by
general values $\tilde{g}_i$ for atom $i$, a straightforward
calculation (see Methods) shows that
\begin{eqnarray}
\langle\dot{J}_z(0)\rangle_c&=&
-\gamma\Big(\sum_{i=1}^{N/2}
  |\tilde{g}_i-\tilde{g}_{i+N/2}|^2|b_i|^2\nonumber\\
&&+\sum_{\stackrel{i,j=1}{i\ne
      j}}^{N/2}\left(\tilde{g}_i^*-\tilde{g}_{i+N/2}^*\right)\left(\tilde{g}_j-\tilde{g}_{j+N/2}\right)b_i^*b_ja_ia_j^*
\Big)\,,\label{outdfs}
\end{eqnarray}
which is in general of order $N^2$. In particular, if the $\tilde{G}_I$ undergo
collective changes $\tilde{G}_I^{(0)}\rightarrow \tilde{G}_I$ with $\tilde{G}_1\ne \tilde{G}_2$ and
if all pairs of atoms
were prepared in the same initial state $a_l=a$, $b_l=b$ for $l=1,2,\ldots, N/2$, we have
\begin{eqnarray}
\langle \dot{J}_z(0)\rangle_c&=&-\gamma
  |\delta\tilde{G}|^2f(N,b)\\
f(N,b)&=&
4|b|^2\Big[\frac{N}{2}+\frac{N}{2}\left(\frac{N}{2}-1\right)(1-|b|^2)\Big]\,\label{outdfs2}\,,
\end{eqnarray}
where $\delta \tilde{G}\equiv (\tilde{G}_1-\tilde{G}_2)/2$. The term
quadratic in $N$ is maximized for $|b|=1/\sqrt{2}$, i.e.~an equal
weight superposition of the two DF basis states $|t_-\rangle_l$ and
$|s\rangle_l$ for each pair of atoms, and gives for $N\gg 1$ a
signal  $\langle n_{\rm ph}\rangle\simeq \gamma\Delta
t|\delta\tilde{G}|^2N^2/4$. As long as the two lattices are not
situated at anti-nodes of the mode, the relation between $\delta
\tilde{G}$ and $\delta L/L$ is linear to lowest order.  If we choose
$x_2-x_1=m \lambda$  with $n_x-1\ge m\in \mathbb{N}$ we
have
\begin{equation}
\delta \tilde{G}=m \pi \cot(\frac{n_x \pi x_1}{L})\frac{\delta
L}{L}\,,
\end{equation}
where we see that $\delta\tilde{G}$ and $\delta L/L$ are related by
a factor independent of $N$.  Note that the measurement of $\langle
n_{\rm ph}\rangle$ allows the measurement of $\delta L/L$, and not
just a detection of a change of $L$~: $\delta L/L=\Big(\langle
n_{\rm ph}\rangle/(\gamma \Delta t(m \pi\cot(n_x\pi
x_1/L))^2f(N,b))\Big)^{1/2}$. The ultimate sensitivity achievable
depends not only on the scaling of the signal $\langle n_{\rm
ph}\rangle$ with $N$, but also of the noise, quantified through the
standard deviation $\sigma(n_{\rm
  ph})=(\langle n_{\rm ph}^2\rangle - \langle n_{\rm ph}\rangle^2)^{1/2}$.
The most fundamental noise associated with the measurement
of $n_{\rm ph}$ is its fluctuation due to the quantum mechanical nature of
the prepared state.  Our approach of
calculating initial time-derivatives of observables by tracing them
over with the Lindbladian implies
$\langle n_{\rm ph}\rangle\ll 1$, as $J_-$ can change the number of
excitations by at most 1, and this condition sets an upper bound on
$\Delta t$. In  this  
regime, $\langle n_{\rm ph}^2\rangle\simeq\langle n_{\rm ph}\rangle$, and,
therefore,
$\sigma(n_{\rm ph})\simeq\sqrt{\langle n_{\rm ph}\rangle}$. It follows that the
signal-to-noise ratio is given by
$\langle n_{\rm ph}\rangle/\sigma(n_{\rm ph})\simeq
|\delta\tilde{G}| \sqrt{\gamma \Delta t}\sqrt{f(N,b)}$. The ultimate
sensitivity achievable can be estimated from a fixed  $\langle n_{\rm
  ph}\rangle/\sigma(n_{\rm ph})$  of order 1, independent of $N$, which leads
to a
minimal $|\delta\tilde{G}|={\cal O}(1/(\sqrt{\gamma \Delta t} N))$. We have
thus shown that a precision measurement based on the purely dissipative
dynamics (\ref{Lc})  and an initial product state
can achieve the Heisenberg limit. This is in contrast to
unitary dynamics of $N$ independent quantum resources, where the SQL cannot
be surpassed when using
an initial product state \cite{Giovannetti06}.

In a real experiment there may be additional fundamental noise sources. One
obvious concern is spontaneous
emission. It is easily verified that
${\cal L}_s[\rho_0]$ leads to a contribution $\langle
\dot{J}_z(0)\rangle_s=-\Gamma \sum_{i=1}^{N/2}|b_i|^2$ to $\langle
\dot{J}_z(0)\rangle$ which scales as
${\cal O}(N)$ and leads to a background signal against which, one might think,
the collective decoherence signal $\langle
\dot{J}_z(0)\rangle_c$ has to be compared. However, note that
spontaneous emission sends photons into the entire open space but {\em not}
into the cavity, whereas the collective emission escapes {\em exclusively}
through the leaky cavity mirror.  Therefore, the two contributions
can be well separated by observing only the photons
escaping through the cavity mirror.


Another obvious concern are fluctuations of the coupling constants. In order
to measure $\langle n_{\rm ph}\rangle$, the experiment has
to be repeated $\nu$ times with $\nu\gg 1$. However, $\nu$ is independent of
$N$ and only given by the desired signal/noise, or, equivalently, by
$\langle n_{\rm ph}\rangle$ itself. Increasing $\nu$ does therefore not
influence the scaling with $N$.  It increases the sensitivity by a factor
$\sqrt{\nu}$, but reduces the bandwidth by $1/\nu$ in the standard way.
While we assume that the time
scale of the mirror
motion is sufficiently  long compared to the time needed for averaging,
the exact coupling constants might fluctuate about
their slowly evolving mean values during the averaging, e.g.~due to
fluctuating traps caused by
vibrations in the set up. But even for perfectly stable traps, the
micro--motion of the
atoms in their respective trapping
potentials, thermal motion, or even quantum fluctuations in the traps will
lead to fluctuating $g_i$. The
cost in sensitivity of these fluctuations depends on their {\em
  correlations}. To see this, let us consider fluctuations
$\delta\tilde{g}_i$ of the
$\tilde{g}_i$ about their mean values $\tilde{G}_I$,
$\tilde{g}_i=\tilde{G}_1+\delta\tilde{g}_i$ for $i=1,\ldots,N/2$,
$\tilde{g}_i=\tilde{G}_2+\delta\tilde{g}_i$ for $
i=N/2+1,\ldots,N$. We
introduce the correlation matrix $C_{ij}=\overline{\delta
  \tilde{g}_i^*\delta \tilde{g}_j}$, where the over-line denotes an average
over the ensemble describing the fluctuations.
Equation~(\ref{outdfs}) then leads to a background $\alpha_{\rm bg}$
in the photon counting rate and to fluctuations $\delta \alpha_f$ on
top of $\alpha_c\equiv -\langle \dot{J}_z(0)\rangle_c$, i.e.\
$\alpha=\alpha_c+\alpha_{\rm bg}+\delta \alpha_f$. The average
background $\overline{\alpha}_{\rm bg}$, given by
\begin{eqnarray}
\overline{\alpha}_{\rm bg}&=&\gamma\Big\{\sum_{i=1}^{N/2}\Big(
C_{ii}+ C_{i+\frac{N}{2}\,i+\frac{N}{2}} -C_{i\,i+\frac{N}{2}}
-C_{i+\frac{N}{2}\,i} \Big)  |b_i|^2\Big\}\nonumber\,,
\end{eqnarray}
can be determined independently at $\delta\tilde{G}=0$, and
subtracted from the signal; it does therefore not influence the
sensitivity of the measurement. The remaining noise $\delta\alpha_f$
fluctuates about zero,
\begin{eqnarray}
\delta\alpha_f&=&2\gamma\Big\{
\sum_{i=1}^{N/2}(\delta\tilde{G}^*\Delta
\tilde{g}_i+\delta\tilde{G}\Delta \tilde{g}_i^*)|b_i|^2+
\sum_{\stackrel{i,j=1}{i\ne j}}^{N/2}(\delta\tilde{G}^*\Delta
\tilde{g}_j+\delta\tilde{G}\Delta \tilde{g}_i^*)b_i^*b_j
a_ia_j^*\Big\}\,,\label{flu}
\end{eqnarray}
where $\Delta\tilde{g}_i\equiv \delta\tilde{g}_i-\delta\tilde{g}_{i+N/2}$.
Assuming real coupling constants, we find the standard deviation
$(\overline{\delta
\alpha_f^2})^{1/2}=4\gamma|\delta\tilde{G}|K$,
where
$K=(\sum_{i,j=1}^{N/2}(C_{i,j}+C_{i+N/2,j+N/2}-C_{i,j+N/2}-
C_{i+N/2,j})S_iS_j)^{1/2}$,
and $S_i=|b_i|^2+b_i^*a_i\sum_{j|j\ne i}b_ja_j^*$ is in general of order
${\cal O}(N)$. This leads to fluctuations in the measured photon number with a
standard deviation $(\overline{\delta n_{\rm ph}^2})^{1/2}$, and to a
signal-to-noise ratio $\langle n_{\rm ph}\rangle/(\overline{\delta n_{\rm
    ph}^2})^{1/2}=|\delta\tilde{G}|f(N,b)/(4K)$.

Several interesting cases can be considered:
\begin{enumerate}
  \item Fully uncorrelated fluctuations, $C_{ij}=C_i\delta_{ij}$, where
  $\delta_{ij}$ stands for the Kronecker-delta: Here we get
  $K=\left(\sum_{i=1}^{N/2}(C_i+C_{i+N/2})S_i^2\right)^{1/2}$, which is in
  general of order $N^{3/2}$, and leads back to the SQL.

\item Pairwise identical fluctuations between the two sets:
  $C_{ij}=C_{i+\frac{N}{2}\,j}=C_{i
  \,j+\frac{N}{2}}=C_{i+\frac{N}{2}\,j+\frac{N}{2}}$ for $
  i,j=1,\ldots,N/2$. This can be the consequence of fully correlated
  fluctucations, $C_{ij}=C \,\,\,\forall\,\,\, i,j$.  Alternatively, such a
  situation arises for
  example for atoms initially
  arranged symmetrically with respect to an anti-node such
  that $\tilde{G}_1^{(0)}=\tilde{G}_2^{(0)}$, if
  the two atoms
  (or ions) in each pair $l$ ($l=1,\ldots,N/2$) are locked into a common
  oscillation. This should be the case for two trapped ions repelling each
  other through a strong Coulomb interaction, and cooled below the
  temperature corresponding to the frequency
  of the breathing mode. Equation~(\ref{flu}) then
  gives $\delta\alpha_f=0$. Note, however, that for initial
  $\tilde{G}_1^{(0)}\ne    \tilde{G}_2^{(0)}$
  the more general DFS leads to a more complicated condition for the
  correlations,
  $C_{ij}|\tilde{G}_2^{(0)}|^2
+C_{i+\frac{N}{2}\,j+\frac{N}{2}}|\tilde{G}_1^{(0)}|^2
-C_{i\,j+\frac{N}{2}}\tilde{G}_2^{(0)*}\tilde{G}_1^{(0)}
-C_{i+\frac{N}{2}\,j}\tilde{G}_1^{(0)*}\tilde{G}_2^{(0)}=0$, which might be
  harder to achieve.
\item Correlated fluctuations within a set, but uncorrelated between the two
  sets, $C_{ij}=C$ for $i,j\in S_1$ or $i,j\in S_2$, but $C_{ij}=0$ for
  $i\in S_1$ and $j\in S_2$ or vice versa. In this case both sums in
  (\ref{flu}) survive and lead to a noise of order ${\cal O}(N^2)$, the
  worst case scenario. However, this comes to no surprise, as such
  correlations are indistinguishable from the signal: all the atoms in a
  given set move in a correlated fashion, but independently from the atoms of the other
  set. This leads to a collective difference in the couplings,
  just as if the length of the cavity was changed.
\end{enumerate}

Case (2) above is clearly the most favorable situation. If there are no
other background signals depending on $N$, we keep the $1/N$ scaling of
$\delta\tilde{G}$ characteristic of the Heisenberg limit.
In order to favor case (2) over cases (1),(3), it appears to be
advantageous to work with ions and to try to
bring the ions in a pair as closely together as possible,
thus strongly correlating their fluctuations, while separating the ions in the
same set as far as possible.

To summarize, we have shown for a particular example how the very sensitive
dependence of collective
decoherence on system parameters can be exploited to reach
the Heisenberg limit in precision measurements while using an initial
product state --- something which is known to be impossible with unitary
dynamics \cite{Giovannetti06}.
It should be clear that the principle of DEM is far more general than the
example exposed here.  Decoherence is itself a
process in which interference effects play an important role. This is
exemplified by the very existence of DFS,  and can
lead to exquisite sensitivity. One might therefore as well try to
exploit these effects instead
of trying to suppress decoherence at all costs.

{\em\bf Methods\\}
Derivation of Eq.~(\ref{outdfs}): The commutation relation $[J_\pm,J_z]=\mp
J_\pm$, valid for any choice of couplings $\tilde{g}_i$,
allows to rewrite $\langle\dot{J}_z(0)\rangle_c=-2\gamma\langle
\psi_0|J_+J_-|\psi_0\rangle$. A short calculation yields
$J_-|\psi_0\rangle=-(1/\sqrt{2})\sum_{i=1}^{N/2}(\tilde{g}_i-\tilde{g}_{i+\frac{N}{2}})b_i|t_-\rangle_i\bigotimes_{l\ne
  i}^{N/2}|\varphi_l\rangle_l$ and leads immediately to
Eq.~(\ref{outdfs}).
Preparation of initial state: In order to prepare the product state
(\ref{psi0}) it is helpful to use three--level atoms with a lambda
structure.  Let $|0\rangle$ and the additional state $|2\rangle$ be
hyperfine (HF) states, and assume that their energies are split in a
sufficiently strong magnetic field, such that only the transition
$|0\rangle\leftrightarrow
|1\rangle$ resonates with the cavity mode. We assume further that the second
optical lattice can be moved along the cavity axis, such that controlled
pairwise collisions of corresponding
atoms in the two lattices can be induced.  Entangled pairs of atoms in their
HF split ground states can thus
be created (for atoms in the same lattice
this has been demonstrated experimentally, see \cite{Bloch08} for a
review). After the creation of an entangled HF state $|\psi'_0\rangle$,
that differs from (\ref{psi0}) by the replacement of states
$|1\rangle$ by states $|2\rangle$, the second lattice is moved back to its
original position.  Now one can
selectively excite the $|2\rangle$ states by a
laser pulse in resonance with the $|2\rangle\leftrightarrow
|1\rangle$ transition, that replaces the singlets in the (very long
lived) HF states by the desired
singlets of
the $|0\rangle$ and $|1\rangle$ states and thus produce the product state
(\ref{psi0}). However, as such, the method is not of much
practical use yet, as it will be virtually impossible to park the second
lattice at the exact position corresponding to coupling constants which
render the state (\ref{psi0}) decoherence free. The extreme
sensitivity of the collective decoherence with respect to changes of the
coupling constants plays against us here, and will lead to a
superradiant flash of light from the cavity after the excitation
$|\psi'_0\rangle\to |\psi_0\rangle$, if the exact position
corresponding to $|\psi_0\rangle\in$ DFS is not achieved. But it is
possible to
position the second lattice at the required position with a precision of
${\cal O}(1/N)$ for the case (2) considered above,
using a feed-back
mechanism and a part of the quantum ressources. With the atoms in the state
$|\psi'_0\rangle$, do the following repeatedly in order to find the
optimal position: Excite a part of the entangled HF pairs containing ${\cal
  O}(N/\ln
N)$ atoms with the laser, measure $\langle n_{\rm ph}\rangle$, and use the
measurement results to bracket
the minimum of $\langle n_{\rm ph}\rangle$ as function of the
lattice position. The minimum of $\langle n_{\rm ph}\rangle$ indicates that
the position
corresponding to the DFS is achieved. Using golden section search, the
minimum  can be bracketed to precision $1/N$
in ${\cal O}(\ln N)$
moves, as at each step the sensitivity of the measurement of the
position of the
lattice is of order ${\cal O}(\ln N/N)\sim {\cal O}(1/N)$. Once the minimum
is found, excite the
remaining unused pairs (there
should be still a number of pairs of ${\cal O}(N)$) to the desired state
$|\psi_0\rangle$.  That state is now decoherence-free, and the system ready
to detect small changes of the position of one of the mirrors. Note that for
this method it is not necessary to know which exact state is produced
in the controlled collisions and subsequent laser excitation.

Imperfections in preparation of $|\psi_0\rangle$: Suppose that instead of
the state (\ref{psi0}) a state
\begin{equation} \label{psi1}
|\tilde{\psi}_0\rangle=\bigotimes_{l=1}^{N/2}|\varphi\rangle_l\mbox{
with }
|\varphi\rangle_l=a|t_-\rangle_l+b|s\rangle_l+c|t_0\rangle_l+d|t_+\rangle_l
\end{equation}
was prepared (we consider the same state for all pairs for
simplicity, but this is not essential). Repeating the calculation
that leads to Eq.~(\ref{outdfs2}) and assuming real $\tilde{G}_i$, we now find
\begin{eqnarray}
\langle n_{\rm ph}\rangle &=&\gamma\Delta t\Big[ \frac{N}{2}
\Big(\big((c-b)\tilde{G}_1+(c+b)\tilde{G}_2\big)^2+d^2(\tilde{G}_1^2+\tilde{G}_2^2)\Big)\nonumber\\
&&+\frac{N}{2}\left(\frac{N}{2}-1\right)
\Big((\tilde{G}_2-\tilde{G}_1)b(a-d)+(\tilde{G}_2+\tilde{G}_1)c(a+d)\Big)^2
\Big]\,.
\end{eqnarray}
The derivative of $\langle n_{\rm ph}\rangle$ with respect to $\tilde{G}_2$
is of order
${\cal O}(N^2)$, and thus still allows to find the minimum of $\alpha$ as
function of
the position of the second lattice with a precision of order ${\cal
  O}(1/N)$ in case 2. At the minimum a
component outside the DFS persists, such that photons will leak out of the
cavity, but the average background is only of order ${\cal
  O}(N)$, and can be measured separately and subtracted from the
singal. Changes $\delta\tilde{G}_2$ of  $\tilde{G}_2$ away from
the position of the minimum still lead to a signal that
scales, for large $N$, quadratically with $N$,
$\langle n_{\rm
  ph}\rangle=\frac{\gamma\Delta
  t}{4}\left(a(c+b)+d(c-b)\right)^2N^2\delta\tilde{G}_2^2$, and the previous
analysis leading to the Heisenberg limit still applies.

\begin{figure}
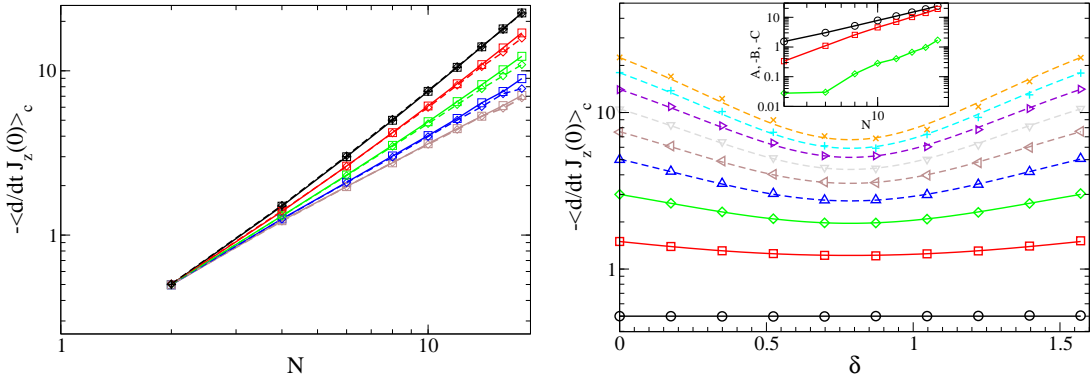

\epsfig{file=scalingZdotSSEV4N-18nop.eps,width=7cm,angle=0}\hspace{0.3cm}
\epsfig{file=zdotofdelta_nop.eps,width=7cm,angle=0}\hspace{0.3cm}
\caption{(a) Scaling of collective photon emission rate $\alpha=-\langle
\dot{J}_z(0)\rangle_c$ (in
units
  $\gamma|\tilde{G}_1-\tilde{G}_2|^2$) in
the mixture of DFS states reached by relaxation from state
$|\tilde{\psi}_0\rangle$, Eq.~(\ref{psi1}). Data for
$\delta=\delta_0$ ($\delta=\pi/2-\delta_0$) denoted by squares and
full lines (diamonds and dashed lines), respectively;
$\delta_0=0,\pi/9,2\pi/9,3\pi/9$ and $4\pi/9$ in black, red, green,
blue, and brown.  Exact analytical results for $\delta=0$,
Eq.~(\ref{outdfs2}) shown with black crosses.  Full and dashed lines
are guides to the eye only. (b) Dependence of $\alpha$ on $\delta$.
$N=2$, 4,6,8,10,12,14,16,18 in black circles, red squares, green
diamonds, blue triangles up, brown triangles left, grey triangles
down, violet triangles right, cyan pluses, orange Xs,
respectively. The full lines for $N$=2,4,6 are exact analytical
results. The dashed lines for $N=8,\ldots,18$ are fits to
$A+B\sin(2\delta)+C\cos(4\delta)$. The inset shows the scaling of
the coefficients $A$ (black circles), $-B$ (red squares), and $-C$ (green diamonds) as function of $N$.}\label{fig.zdot}
\end{figure}

Alternatively, one can get rid of the additional background by letting
the system relax before measuring changes of $L$. Indeed, any state
with a component outside the DFS will relax to a DFS state (and thus
a dark state) or mixtures of DFS states within a time of order
$1/\gamma$ or less. Components with large total pseudo-angular
momentum $J$ relax in fact in much shorter time of order $1/(J\gamma)$.
The   DFS states reached through relaxation starting from
$|\tilde{\psi}_0\rangle$ still allow a scaling of $\langle n_{\rm
  ph}\rangle$ close to
$N^2$. We have shown this by simulating the relaxation process with
the help of the stochastic Schr\"odinger equation (SSE)
corresponding to Eq.~(\ref{Lc}).  For real values of the coupling
constants, the SSE reads
\begin{eqnarray}
d\psi(t)&=&D_1(\psi(t))\,dt+D_2(\psi(t))\,dW(t)\\
D_1(\psi)&=&\gamma\left(2\langle J_-\rangle_\psi J_--J_+J_--\langle
J_-\rangle^2_\psi\right)\psi\\
D_2(\psi)&=&\sqrt{2\gamma}\left(J_--\langle J_-\rangle_\psi\right)\psi\,,
\end{eqnarray}
where $dW(t)$ is a Wiener process with average zero and variance
$dt$, and $\langle J_-\rangle_\psi=\langle \psi|J_-|\psi\rangle$
\cite{Breuer06}. Using an Euler scheme with a time step of
$0.01/\gamma$, we followed the convergence of $\psi(t)$ to DFS
states for states with
$(a,b,c,d)=(\cos\delta,1,0,\sin\delta)/\sqrt{2}$, until the norm of
the difference $|\psi(t+dt)\rangle-|\psi(t)\rangle$ dropped below
$10^{-12}$. In these final dark states, randomly distributed over
the DFS, we calculated $\alpha$, and averaged over a large number
$n_r$ of realizations of the stochastic process ($n_r=10^5$, $10^4$,
$10^4$, $2.5\cdot 10^3$, $10^3$, $1.25\cdot 10^3$, 250, 200, and 250
for $N=2,4,6,8,10,12,14,16$, and $18$).  Figure \ref{fig.zdot}
shows the scaling of $\alpha$ as function of $N$ for different
values of $\delta$ for $0\le\delta\le\pi/2$ up to $N=18$. Within
this numerically accessible range of $N$, $\alpha$ follows a power
law $\alpha\propto N^p$ with an exponent $p$ that decays only
gradually with $\delta$ for $\delta\le \pi/4$. Moreover, that decay
might be a finite size effect: Note that, surprisingly,
$\alpha(\delta)$ appears to be close to symmetric with respect to
$\delta=\pi/4$.  This is corroborated by exact analytical
calculations based on the diagonalization of ${\cal L}_c$, which
lead to $\alpha=1/2$ for $N=2$,
$\alpha=(55-12\sin(2\delta)-\cos(4\delta))/36$ for $N=4$, and
$\alpha=(303-110\sin(2\delta)-3\cos(4\delta))/100$ for $N=6$ (in
units $\gamma|\tilde{G}_1-\tilde{G}_2|^2$). The plot shows that all
numerical data can be very well fitted by
$\alpha=A+B\sin(2\delta)+C\cos(4\delta)$. From Eq.~(\ref{outdfs2})
we know that $A+C$ has to scale as $N^2$ for sufficiently large $N$.
Both $B$ and $C$ are negative for all $N$ for which we have data,
and $C$ appears to be negligible. Fig.~\ref{fig.zdot} shows that
$-B$ increases even more rapidly than $N^2$ (a fit in the range
$N=8,\ldots,18$ gives a power law $N^{2.4}$). But $B$ has to cross
over to a power law $N^p$ with $p\le 2$, unless other Fourier
components start contributing significantly. Otherwise, $\alpha$
would become negative for $\delta>0$. This indicates that for large
$N$ the scaling of $\alpha$ is in fact $N^2$ for all $\delta$.

In summary, our
method still
works, even if the product state (\ref{psi0}) is not prepared
perfectly. One has the choice to start measurement immediately after state
preparation, which gives an additional background of order $N$, or to wait a
time of the
order of a
few $1/\gamma$ after preparation of the initial state, until no more
photons leave the cavity through the mirror, with no additional
background. In both cases the scaling of the signal-to-noise ratio is still
$\sim \sqrt{\gamma\Delta t}|\delta L/L| N$, and allows to reach the
Heisenberg limit in the measurement of a subsequent small change of $L$.

Another class of states in the DFS that allows quadratic scaling of
$\alpha$ with
$N$, are Schr\"odinger cat states of
macroscopic pseudo-angular momentum $j_1=j_2\equiv \ell\sim N$ in the two
sublattices (i.e.~states $|(\ell,\ell)j,-j\rangle$),
\begin{align}\label{mv}
    \langle(\ell,\ell) j,-j|J_{1+}J_{1-}|(\ell,\ell)j,-j\rangle=
    {}&|\tilde{G}_1|^2
    \frac{(2j+1)(2j+2)}{4(j+1)^2-1}\left[\ell(\ell+1)-\frac{j(j+2)}{4}\right]\,,
\end{align}
with $J_{1-}=\tilde{G}_1\sum_{i=1}^{N/2}\sigma^{(i)}_-$
in agreement with the initial
intuitive reasoning.
If the total
system is initially in a singlet state ($j=0$) and the angular
momentum of each of the two sets of $N/2$ atoms has its maximal
value $\ell=N/4$, we have
\begin{equation}\label{idcq}
   \langle n_{\rm ph}\rangle=\frac{\gamma\Delta t}{3}|\delta\tilde{G}|^2\,
   N(N+4)\,.
\end{equation}
While these states are protected by the DFS and thus do not suffer the fate
of rapid decoherence of the highly entangled states proposed for QEM, it
appears to be still more challenging to produce
them compared to the product states (\ref{psi0}). Our numerical simulations
also show that  states
chosen randomly inside the DFS lead on the average only to $\langle n_{\rm ph}\rangle\propto
N$. Therefore, it is rather remarkable that the product states
(\ref{psi0})
and the above
Schr\"odinger cat states share the property of scaling of $\langle n_{\rm
  ph}\rangle\propto N^2$.

The authors declare to have no competing financial interests.

{\em Acknowledgments:} DB thanks Eite Tiesinga and Peter Braun for useful
discussions. This work was supported by the Agence
National de la Recherche (ANR), project INFOSYSQQ. Numerical calculations
were partly performed at CALMIP, Toulouse. J.M.\ thanks the Belgian
F.R.S.-FNRS for financial support.
\bibliography{../mybibs_bt}

\end{document}